\documentclass[preprint2, longabstract]{aastex}
\shorttitle{Whole Earth Telescope observations of the pulsating subdwarf B star PG~0014+067}
\shortauthors{Vu\v{c}kovi\'{c} et al.}

\begin{document}

\title{Whole Earth Telescope observations of the pulsating subdwarf B star PG~0014+067}

\author{M. Vu\v{c}kovi\'{c}\altaffilmark{1,2}, S.D. Kawaler \altaffilmark{1}, 
S. O'Toole\altaffilmark{3}, Z. Csubry\altaffilmark{4}, 
A. Baran\altaffilmark{5}, S. Zola\altaffilmark{5,6}, 
P. Moskalik\altaffilmark{7}, 
E.W. Klumpe\altaffilmark{8}, R. Riddle\altaffilmark{9}, 
M.S. O'Brien\altaffilmark{10}, F. Mullally\altaffilmark{11}, 
M.A. Wood\altaffilmark{12}, V. Wilkat\altaffilmark{12},  
A.-Y. Zhou\altaffilmark{13}, M.D. Reed\altaffilmark{13}, 
D.M. Terndrup\altaffilmark{14}, D.J. Sullivan\altaffilmark{15}, 
S.-L. Kim \altaffilmark{16}, W.P. Chen\altaffilmark{17}, 
C.-W. Chen\altaffilmark{17}, W.-S. Hsiao\altaffilmark{17}, 
K. Sanchawala\altaffilmark{17}, H.-T. Lee\altaffilmark{17}, 
X.J. Jiang\altaffilmark{18}, R. Janulis\altaffilmark{19, 20}, 
M. Siwak\altaffilmark{6}, W. Ogloza\altaffilmark{5}, 
M. Papar\'{o}\altaffilmark{4}, Zs. Bogn\'{a}r\altaffilmark{4}, 
\'{A}. S\'{o}dor\altaffilmark{4}, G. Handler\altaffilmark{21}, 
D. Lorenz\altaffilmark{21}, B. Steininger\altaffilmark{21}, 
R. Silvotti\altaffilmark{22}, G. Vauclair \altaffilmark{23}, 
R. Oreiro\altaffilmark{24}, R. \O stensen\altaffilmark{25}, 
A. Bronowska\altaffilmark{25}, B.G. Castanheira\altaffilmark{26}, 
S.O. Kepler\altaffilmark{26},  L. Fraga\altaffilmark{27}
H.L. Shipman\altaffilmark{28}, J.L. Provencal\altaffilmark{28}, 
D. Childers\altaffilmark{28} }

\altaffiltext{1}{Department of Physics and Astronomy, Iowa State University, Ames, IA 50011 USA.}
\altaffiltext{2}{Instituut voor Sterrenkunde, Celestijnenlaan 200B, 3001 Leuven, Belgium.}
\altaffiltext{3}{Dr. Remeis-Sternwarte, Astronomisches Institut der Universit\"{a}t Erlangen-N\"{u}rnberg, Sternwartstr. 7, Bamberg 96049, Germany.}
\altaffiltext{4}{Konkoly Observatory, P.O.Box 67, H-1525 Budapest XII, Hungary.}
\altaffiltext{5}{Mt. Suhora Observatory, Cracow Pedagogical University, Ul. Podchorazych 2, PL-30-084 Cracow, Poland.}
\altaffiltext{6}{Astronomical Observatory, Jagiellonian University, Ul. Orla 171, 30-244, Cracow, Poland.}
\altaffiltext{7}{Nicolas Copernicus Astronomical Center, Polish Academy of Sciences, Ul. Bartycka 18, 00-716 Warsaw, Poland.}
\altaffiltext{8}{Department of Physics and Astronomy, Middle Tennessee State University, Murfreesboro, TN 37132 USA.}
\altaffiltext{9}{California Institute of Technology, Pasadena, CA 91125 USA.}
\altaffiltext{10}{Department of Astronomy, Yale University, New Haven, CT 06851, USA.}
\altaffiltext{11}{Department of Astronomy, University of Texas, Austin, TX 78712, USA.}
\altaffiltext{12}{Department of Physics and Space Sciences \& SARA Observatory, Florida Institute of Technology, Melbourne, FL 32901, USA.}
\altaffiltext{13}{Department of Physics, Astronomy and Material Science,  Missouri State University, Sprinfield, MO 65897, USA.}
\altaffiltext{14}{Department of Astronomy, The Ohio State University, Columbus, OH 43210, USA.}
\altaffiltext{15}{Department of Physics, Victoria University of Wellington,P. O. Box 600, Wellington, New Zealand.}
\altaffiltext{16}{Korea Astronomy and Space science Institute, Daejeon, 305-348, Korea (South).}
\altaffiltext{17}{Institute of Astronomy, National Central University Taiwan, 32054 ChungLi, Taiwan.}
\altaffiltext{18}{National Astronomical Observatories, Chinese Academy of Sciences, Beijing 100012, PR China.}
\altaffiltext{19}{Institute of Theoretical Physics and Astronomy (ITPA), Go\v{s}tauto 12, Vilnius 2600, Lithuania.}
\altaffiltext{20}{Astronomical Observatory of Vilnius University, \v{C}iurlionio 29, Vilnius 2009, Lithuania.}
\altaffiltext{21}{Institut f\"{u}r Astronomie, T\"{u}rkenschanzstr. 17, A-1180 Wien, Austria.}
\altaffiltext{22}{INAF - Osservatorio Astronomico di Capodimonte, via Moiariello 16, I-80131 Napoli, Italy.}
\altaffiltext{23}{Universit\'{e} Paul Sabatier, Observatoire Midi-Pyr\'{e}n\'{e}es, 14 avenue E. Belin, 31400 Toulouse, France.}
\altaffiltext{24}{Departamento de F\'{i}sica Aplicada, Universidad de Vigo, 36200, Spain.}
\altaffiltext{25}{Isaac Newton Group of Telescopes, Santa Cruz de La Palma 37800, Canary Islands, Spain.}
\altaffiltext{26}{Instituto de F\'{i}sica, Universidade Federal de Rio Grande do Sul,  91501-900 Porto-Alegre, RS, Brazil.}
\altaffiltext{27}{Departamento de F\'{\i}sica, Universidade Federal de Santa
Catarina, Florian\'opolis, Brazil}
\altaffiltext{28}{Department of Physics and Astronomy, University of Delaware, 223 Sharp Laboratory, Newark, DE 19716, USA.}

\begin{abstract}
PG~0014+067 is one of the most promising pulsating subdwarf B stars for seismic analysis, as it has a rich pulsation spectrum. The richness of its pulsations,
however, poses a fundamental challenge to understanding the pulsations of 
these stars, as the mode density is too complex to be explained only with 
radial and nonradial low degree ($\ell<3$) $p$-modes without rotational 
splittings. One proposed solution, suggested by  \citet{Brassard2001} for the 
case of PG~0014+067 in particular, assigns some modes with high degree 
($\ell=3$). On the other hand, theoretical models of sdB stars suggest that 
they may retain rapidly rotating cores \citep{KH2005}, and so the high mode 
density may result from the presence of a few rotationally--split triplet 
($\ell=1$), quintuplet ($\ell=2$) modes, along with radial ($\ell=0$) 
$p$-modes. To examine alternative theoretical models for these stars, we need 
better frequency resolution and denser longitude coverage. Therefore, we 
observed this star with the Whole Earth Telescope for two weeks in October 
2004. In this paper we report the results of 
Whole Earth Telescope observations of the pulsating subdwarf B 
star PG~0014+067. We find that the frequencies seen in PG~0014+067 do not 
appear to fit any theoretical model currently available; however, we find a 
simple empirical relation that is able to match all of the well-determined 
frequencies in this star.
\end{abstract}


\section{Introduction}

Even though stellar evolution is one of the most mature areas in theoretical astrophysics, there are still phases in the life of an ordinary star (like our Sun) that are not well understood.  We have a nearly continuous evolutionary path of a low massive star all the way from its birthplace, a protostellar cloud,  to its graveyard, with few gaps in our understanding.  However,
at least one major gap remains: the core helium flash that initiates helium core burning.  Unlike most phases of stellar evolution, the flash itself occurs in a matter of minutes rather than millions of years. It is essentially a massive runaway nuclear reaction near the center of the star that releases a huge amount of energy on a short time scale. Despite ignition on a dynamical time scale, apparently this event does not  make the star explode, because we $do$ see stars that are clearly survivors of this process.   Post helium core flash objects are identified with horizontal branch stars in globular clusters, and their field counterparts.

At the extreme blue end of the horizontal branch, subdwarf B (sdB) stars are survivors of the core helium flash that have a small hydrogen rich envelope surrounding the helium--burning core.  According to  evolutionary calculations, for example those by \citet{DROC93}, sdB stars are core helium burning stars with a canonical mass of $M\approx$ 0.5 $M_{\odot}$ and a very thin, inert hydrogen envelope with $\frac{M_{H}}{M_{*}}  \sim 0.0004 $ which places them on the hot end of the horizontal branch (HB), the so-called extreme horizontal branch (EHB). The fact that sdB stars have lost almost all of their hydrogen layer is what makes them so enigmatic from the stellar evolution point of view. To lose that much mass, they have to suffer a considerable mass loss during the RGB and most probably during the helium core flash. 

Details of the origin of the sdB stars are largely unknown. There are currently many competing theories for their origin which involve several possible channels. These include single star evolution with an extreme mass loss on the RGB \citep{D'Cruz96}, or different binary interactions: common envelope binary evolution \citep{Mengel76}, stable Roche lobe overflow (\citet{Han02, Han03} and references therein), or merging of two helium white dwarfs \citep{Iben90}. 

The most fundamental missing piece to our understanding of the evolution of the sdB stars is the nature and physics behind their mass loss \citep{FPR76}.  To end up on the EHB they must lose nearly all of their hydrogen at almost exactly the same phase as the helium core has attained the minimum mass required for the helium flash to occur.  By exposing the interior properties of
these survivors we can learn how the stars manage to survive this event of unstable helium ignition and plug this gap in the story of stellar evolution. 

Luckily, some of these stars pulsate, letting us therefore use the tools and techniques of asteroseismology to probe their interiors. Discovered by \citet{discoveryK97}, the pulsating sdB stars (officially designated V361 Hya stars, but also called sdBV stars by analogy with the pulsating white dwarfs) are short period, low amplitude multimode pulsators. Typical periods are about 100-250s with an overall period range of 80-600s. The pulsation amplitudes are generally less than a few hundredths of a magnitude. The short periods, being of the order of, and shorter, than the radial fundamental mode for these stars, suggest that the observed modes are  low-order, low-degree $p-$modes \citep{Charpinet2000}. 

The rapidly pulsating sdB stars occupy a region in the $\log{g} - T_{\rm eff} $  plane with effective temperatures of $T_{\rm eff}$ = 29 000 - 36 000 K and surface gravities $\log{g} $ = 5.2 - 6.5.  Many of the high gravity pulsating sdB stars show complex pulsations that require a large number of modes. Radial modes alone are not enough to explain the complex light curves of these stars, therefore nonradial modes must also be present \citep{K99, Kilkenny99, Charpinet2000}. 

Among the sdBV stars PG~0014+067 stands out as one of the richest pulsators. It was found to be an sdBV star in June 1998 \citep{Brassard2001}.  While close binaries like KPD~1930+2752 \citep{KPD2003} and lower-gravity sdB pulsators like PG~1605+072 \citep{Kilkenny99} can show dozens of modes, PG~0014+067 is a higher gravity single sdB star that shows a large number of nonradial modes, with at least sixteen reported in the discovery paper.  Periods range from 80 to 170 seconds.  It is a ``typical'' sdBV star with an effective temperature of $T_{\rm eff} $ = 33 310K and $\log {g} = 5.8 $ (Brassard et al. 2001). The richness of its pulsation, however, makes it a star of vital interest to the study of sdBV stars and, by extension, the entire hot horizontal branch phenomenon. The rich mode density in PG~0014+067 poses a fundamental challenge to understanding the pulsations of these stars, as there are not enough radial and nonradial modes with low degree $\ell$ ($\ell <3$) to account for all of them (without rotational splitting). Low degree nonradial modes have a limited frequency distribution which implies that we need to resort to higher values of $\ell$ to account for all of the observed periodicities.  That presents further challenges, as geometrical cancellation increases rapidly with $\ell$ so that higher $\ell$ modes need to have a much larger intrinsic amplitude than low $\ell$ modes to reach the same observable amplitiude.

One proposed solution, by \citet{Brassard2001}, presents an envelope structural model that approximates their determined pulsation frequencies with $p$-modes of degrees 0, 1, 2, and 3.  Their resulting seismic model provides an estimate of the mass of PG~0014+067 of 0.49 $\pm$ 0.02 $M_{\odot}$ and, in addition, matches the observed spectroscopic parameters, and places strict limits on the thickness of the surface hydrogen layer.

Another possible explanation for the mode density in PG~0014+067 comes from theoretical calculations of the evolution of stars on the red giant branch, and the subsequent structure on the horizontal branch.  These models show that stars can develop a rapidly rotating core on the red giant branch, and then preserve that core on the horizontal branch.  If that is the case, then certain modes of pulsation in sdBV stars could show large rotational splitting - much larger than suggested by the (slow) surface rotation velocity.  Models by \citet{KH2005}, in particular point out that rotational splittings of low order $\ell$ = 1 and $\ell$ = 2 modes can cause sufficient frequency peaks in the observed range provided that the stellar core rotates much faster than its surface.  These models show rotational splitting that can vary by factors of two or more from one mode to the next.  If that is the case in PG~0014+067, then it could sufficiently complicate the observed mode spectrum with modes of $\ell$=0, 1, and 2 alone.

The only published observations of PG~0014+067 are single-site data from \citet{Brassard2001}, who have observed this star using the 3.6 meter Canada-France-Hawaii Telescope (CFHT) during five consecutive nights (1.9h average run length) in June 1998. These data show that the oscillation spectrum is complex, but being single-site data they suffer from severe one cycle per day aliases making the frequency identification rather difficult.  

To fully exploit this star we need to determine, unambiguously, the oscillation frequencies present. This can be achieved only by extended longitude coverage with a sufficient frequency resolution to be able to resolve the real pulsations from the inevitable one cycle per day aliases.  With a secure set of oscillation frequencies, we can critically examine the proposed theoretical models for these stars without concern for fitting artifacts caused by diurnal aliasing instead of true oscillation modes. PG~0014+067 was selected as the primary target for a Whole Earth Telescope (WET) campaign, held in October 2004. 
In this paper, we present the results of this multisite campaign.  In \S2 we present the observations followed by a review of the data reduction procedure. In Section 3, we report on the frequency analysis of the WET data set on PG~0014+067, and we discuss the results in \S4 along with the plans for future work. 
  
\section{Observations}

PG~0014+067 was a primary target during the 24th WET run, Xcov24, held in 2004. Being a faint star, with $V \approx 16.5$, and a low amplitude pulsator - most modes have amplitudes below 2 mmag - we sought larger than average ($\geq$2 meter class) telescopes distributed around the Earth. Time was allocated to five such telescopes ($>$ 2m) that were well-distributed in longitude.  Observations spanned two weeks centered around New Moon from October 7-21, 2004.  With a right ascension near 0h, and a near--equatorial declination of +7 degrees, PG~0014+067 was equally accessible from both Northern and Southern Hemisphere sites during this time of the year.  As a contingency, smaller telescopes ($<$ 2 m) with CCD detectors were included to minimize the gaps in the data and to simplify the alias pattern of our spectral window.  Together with 13 smaller telescopes, 18 sites participated in this campaign. 

Unluckily, non-photometric conditions all over the globe during the main campaign resulted in less-than-ideal coverage. Table 1 lists all the observations gathered during Xcov24 on PG~0014+067.  As can be noticed from the log of the observations given in Table 1, much of the data did not come from our bigger telescopes ($>$ 2m), because of the bad weather conditions, but from moderate size ones (between 1m and 2m). The 1.0 m telescope at Lulin Observatory in Taiwan contributed data on six nights.  An additional four nights of high--quality data came from the Bohyunsan Optical Astronomy Observatory (BOAO) 1.8m telescope in South Korea.  Also, surprisingly good data were obtained by the SARA 0.9m telescope at KPNO, given the relatively small size of the telescope.  Despite the terrible weather and some unexpected instrumental problems we had during Xcov24, the two week core campaign resulted in about 180 hours of data on the primary target, yielding a duty cycle of about 58 \%. 
The WET run on PG~0014+067 spans two weeks which results in a formal frequency resolution of $\Delta \nu= 0.9 \mu$Hz.  

% Table 1
\begin{deluxetable}{llcccc}

\tabletypesize{\scriptsize}

\tablecaption{Journal of observations of PG~0014+067 in October 2004.}

\tablehead{
\colhead{Run name} & \colhead{Telescope [m]} & \colhead{UT Date} &\colhead{Start [UT]} & \colhead{Duration [hrs]} & Instrument
}
\tablewidth{0pt}
 
\startdata
mdr275\tablenotemark{*} & MDM 1.3 & 05 & 08:55:15 & 2.10 & CCD\\
teide-10072004$\ast$ & Tenerife 0.8 & 08 & 02:36:40 & 0.55 & CCD \\
sara213 & SARA 0.9 & 08 & 02:47:59 & 8.74 & CCD\\
PG0014\_08oct04 & Loiano 1.5 & 08 & 19:24:50 & 5.50 & CCD \\
sara214 & SARA 0.9 & 09 & 01:58:59 & 9.20 & CCD\\
PG0014\_09oct04\tablenotemark{*} & Loiano 1.5 & 09 & 19:45:09 & 1.44 & CCD\\
PG0014\_10oct04 & Loiano 1.5 & 10 & 19:09:20 & 6.35 & CCD \\
lulin-0001 & Lulin 1.0 & 11 & 11:10:35 & 8.35 & CCD \\
041012A & NAOC 2.16 & 12 & 12:27:00 & 7.50 & PMT\\
korea-001 & BOAO 1.8 & 12 & 13:02:10 & 5.45 & CCD\\
korea-002 & BOAO 1.8 & 13 & 10:19:20 & 5.32 & CCD\\
041013A\tablenotemark{*} & NAOC 2.16 & 13 & 11:27:22 & 8.50 & PMT\\
lit-01 & Moletai 1.56 & 13 & 21:49:20 & 4.65 & PMT\\
france-01\tablenotemark{*} & Pic Du Midi 2.0 & 13 & 20:50:00 & 2.85 & PMT\\
041014A & NAOC 2.16 & 14 & 11:23:30 & 8.49 & PMT\\
korea-003 & BOAO 1.8 & 14 & 10:55:16 & 7.27 & CCD\\
lit-05 & Moletai 1.56 & 14 & 19:00:10 & 7.37 & PMT\\
korea-004 & BOAO 1.8 & 15 & 10:19:00 & 7.75 & CCD\\
A0934 & McDonald 2.1 & 15 & 06:26:07 & 2.65 & CCD\\
NOT-001 & NOT 2.6 & 15 & 19:55:23 & 4.32 & CCD \\
NOT-002 & NOT 2.6 & 16 & 00:19:52 & 4.87 & CCD\\
lulin-0008 & Lulin 1.0 & 16 & 12:34:26 & 6.45 & CCD \\
NOT-003 & NOT 2.6 & 16 & 19:43:22  & 8.37 & CCD\\
rkb1\tablenotemark{*} & LNA 1.6 & 17 & 02:25:04 & 3.15 & CCD\\
A0939 & McDonald 2.1 & 17 & 05:19:58 & 1.04 & CCD\\
A0940\tablenotemark{*} & McDonald 2.1 & 17 & 07:32:18 & 1.32 & CCD \\
lulin-0010 & Lulin 1.0 & 17 & 11:36:36 & 7.51 & CCD\\
maja10182004 & KPNO 2.1 & 18  & 09:06:40 & 1.78 & CCD \\
lulin-0012 & Lulin  1.0 & 18 & 10:50:02 & 3.10  & CCD\\
041018B & NAOC 2.16 & 18 & 13:47:00 & 5.88 & PMT\\
maja10192004 & KPNO 2.1 & 19 & 05:19:40& 3.40 & CCD\\
lulin-0015 & Lulin 1.0 & 19 & 10:43:11 & 8.33 & CCD \\
maja10202004 & KPNO 2.1 & 20 & 02:11:20 & 3.18 & CCD\\
maja10202004b& KPNO 2.1 & 20 & 09:39:40 & 0.93 & CCD\\
lulin-0016 & Lulin 1.0 & 20 & 11:35:48 & 7.35 & CCD\\
hungary-007\tablenotemark{*}& Konkoly 1.0 & 20 & 18:09:56 & 1.56 & CCD
\enddata
\tablenotetext{*}{not used because of low signal-to-noise}
\end{deluxetable}
The telescopes used for the primary target range in aperture from 0.9m to 2.56m. Data from NAOC (China), Moletai Obs. (Lithuania) and Pic Du Midi (France) were obtained using three-channel PMTs as described by \citet{Kleinman1996}. Exposure times were 5 seconds for 2m telescopes (China and France) and 10 seconds for the 1.65m telescope in Lithuania.  Channel 1 measured the program star, channel 2 measured a local comparison star and the third channel simultaneously recorded sky background. All PMT photometric observations were taken in white light to maximize the photon count rate.

All other data were obtained using CCDs (see column Instrument of Table 1). For the CCD measurements, exposures ranged from 5-15 seconds depending on the size of the telescope. As the readout times differ from chip to chip and were quite long for some sites, observations were made in the ``windowed'' mode, i.e. only a portion of the chip was read out to minimize the cycle time, making sure there were always at least two comparison stars in the frame. Observations from McDonald observatory and LNA were made in the frame transfer mode.
The actual CCD frames at every individual site were never bigger than 5 x 5 arc minutes centered around the target. The number of comparison stars depended on the size of this window, having from two to six comparison stars in the frame. 
In this way, consecutive data points were obtained in 5 to 25 s intervals, depending on the instrument. Data from McDonald and KPNO used BG\,40 filters and data from NOT used a W\#92 filter.
Together with the science frames, each night of observation included calibration images (bias, dark and flat field frames). 

Preliminary, ``on the fly'' data reduction was done during the campaign at a central headquarters (at Iowa State University). Since the WET collaboration does not operate with a standard CCD photometry extraction package, raw CCD frames were pre-reduced (overscan, bias, dark, flat filed corrected) by the observers using their own software immediately after the run,  and the first-look photometry was sent to the headquarters (HQ) in the standard WET/XQED format. Incoming data were further reduced at HQ using the WET software XQED \citep{Reed2003}.  These preliminary reductions were used during the course of the run to allocate telescopes between the primary target and secondary targets, and for preliminary analysis of the pulsation spectrum of PG~0014+067.  For this paper, all the data were carefully re-reduced by the lead author from the original CCD frames.

For the PMT data, reduction followed standard WET procedures. After the proper dead time correction, the sky background (channel three) was subtracted both from the target and from the comparison star point by point. When necessary, the variable star data were then divided by the comparison star to correct for transparency variations. To correct for any low-frequency variations caused by seeing and differential color extinction, the resulting light curve was then fitted with a low order polynomial. Bad points, coming from either electronic noise, guiding errors or clouds, were removed by hand. Finally, the light curve was mean subtracted leaving only the fractional variations from the mean ($ma= \Delta I / I $) and a barycentric correction to the exposure mid-times was applied giving the times of the data points projected to the barycenter of the solar system. 

For our final reduction, the raw CCD data were reduced using the IRAF
CCDPROC package and the reduction comprised correction for bias, dark counts and flat-field. Further photometric measurements (aperture photometry) on these reduced frames were made using the PHOT package. After carefully examining the extracted aperture photometry for a set of different apertures for each observation, the aperture that gave the best signal-to-noise ratio was used. The apertures that typically gave the optimal signal-to-noise ratio in the resulting light curve had a radius of $\approx$1.75 times the FWHM, with the surrounding annulus to determine the local sky level. In this way each star in the frame has its own corresponding sky channel. The aperture photometry measurements derived in this way were then imported into XQED. 
The resulting light curves, showing a fractional variation of the intensity of light as a function of time are presented in Figure 1.

% Figure 1
\begin{figure}
\includegraphics[scale=0.35,angle=-90]{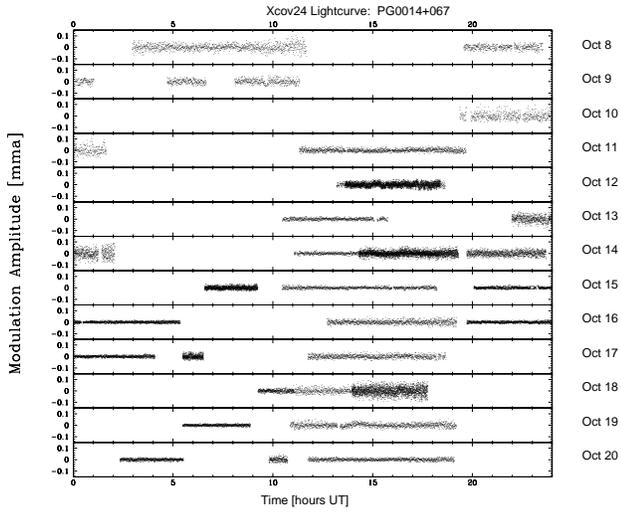}
\caption{Light curve of PG~0014+067 obtained during Xcov24.  Each panel spans 24 hours, 
on the UT date indicated. Different density of data points comes from several factors:  overlapping of the data, different sampling time and different signal-to-noise ratio.}
\end{figure}
\section{The frequencies of PG~0014+067}

Combining all the light curves described in the previous section (Figure 1) 
we computed the Fourier transform (FT) to deduce the periodicities present in the data. To optimize the number of detected intrinsic frequencies present in the total light curve, gathered with the different aperture telescopes (from 0.9m-2.56m), we weighted the data using the weighting scheme favored by Handler (2003).  Weights are determined for individual runs, with the weight for a run taken as inversely proportional to the mean point-to-point scatter within the run. However, to some degree, weighting inflates the amplitude of the daily aliases and widens the peaks. Consequently, weighting may affect the ability to resolve closely spaced peaks. Bearing this in mind weighted data were used to expose and identify periodicities, but the frequencies, amplitudes and phases of the individual peaks were determined using the `unweighted' data. The weighted FT of the full data set on PG~0014+067 gathered during Xcov24 is depicted in Figure 2 showing the `modulation amplitude', in units of mma= ma/1000, where ma=$ \Delta {I}/I$, of variations in the detected intensity $I$ at a given frequency. 

% Figure 2
\begin{figure}
%\epsscale{0.80}
\plotone{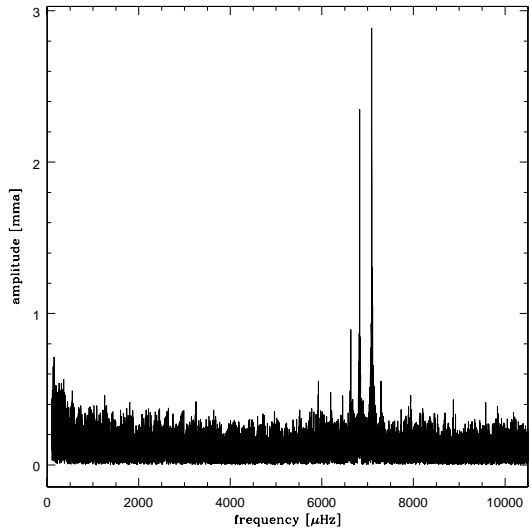}
\caption{Weighted Fourier transform (amplitude spectrum) of the complete WET data set on PG~0014+067.}
\end{figure}
To see the effect weighting has on the amplitude spectrum we have calculated the spectral window for the weighted and unweighted data.  Figure 3 shows the spectral window of the whole WET run on PG~0014+067 in both cases.  The central peak in the plot corresponds to the input frequency of a single sinusoid; all other peaks are aliases caused by the gaps in the data set.

% Figure 3 
\begin{figure}
%\plotone{f3.eps}
\includegraphics[scale=0.30,angle=-90]{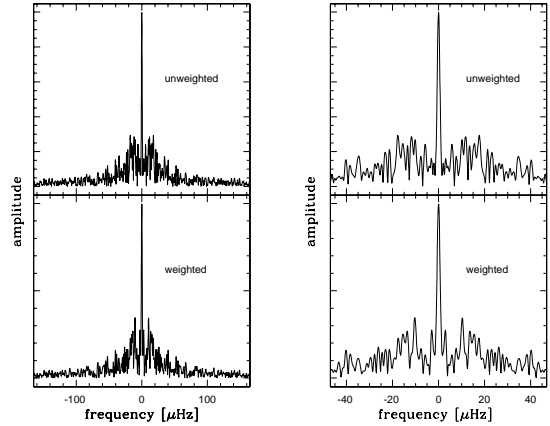}
\caption{Spectral window for the PG~0014+067 WET data set.  The expanded frequency range on the left displays the impact of the data sampling on the noise well beyond the 1 $d^{-1}$, while the closer view on the right shows the details of the alias pattern.  Both cases (top and bottom) sample a noise-free sinusiod; the weights for the data were applied to the sinusoid to compute the window patterns shown in the bottom panels.  The vertical amplitude scale is amplitude (in arbitrary units).}
\end{figure}

\subsection{Frequencies found in the Xcov24 data}

As can be clearly seen in Figure 2, the main action is concentrated in a very narrow frequency range. It appears as if all the power is distributed into two to three modes around the 7000$\mu$Hz with maximum intensity variations of $\sim$ 3 mma. This is a typical signature of a low amplitude sdB pulsator \citep{Kilkenny2002}. Figure 2 confirms that there are no frequencies with amplitudes above the mean noise level at higher frequencies, up to the Nyquist frequency ($f_{N}\approx 30 000\mu$Hz). Therefore, we can securely focus our analysis on the narrow region of the main power. The spectral region above  $20 000\mu$Hz is not plotted as it contains no additional information.
 
% Figure 4 
\begin{figure}
\plotone{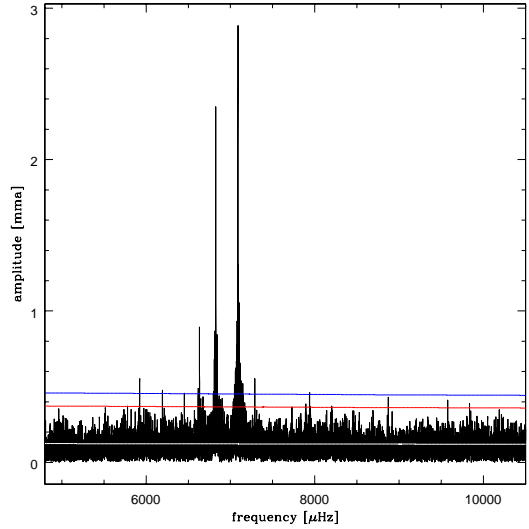}
\caption{Weighted FT of the PG~0014+067 WET data expanded to the region of the most power.  The mean noise level ($\sigma_{\rm noise}$), 3$\sigma_{\rm noise}$, and 3.7$\sigma_{\rm noise}$ are shown as nearly horizontal lines (from bottom to top respectively).}
\end{figure}

Figure 4 expands the FT of the entire data set around the region of the main power, 
$5000-10000\mu$Hz.  The solid white line in Figure 4 corresponds to the noise level, $\sigma_{\rm noise}$, the corresponding red (dark) line represents 3 times the noise level, $3\sigma_{\rm noise}$, while the blue (top) line represents 3.7 times the noise level, $3.7\sigma_{\rm noise}$.  The noise in the power spectrum is calculated by averaging the amplitudes around each frequency in a wide frequency range.
For our criterion determining which peaks are ``real'' signals in the power spectrum, we adopt a conservative significance threshold at $3.7\sigma_{\rm noise}$. \citet{Kuschnig1997} estimate that this results in a 99 per cent confidence limit. The $3\sigma_{\rm noise}$ threshold is adopted by others as sufficient \citep{Brassard2001, Metcalfe2005, Charp2005} which would correspond to 80 per cent confidence limit \citep{Kuschnig1997}. Hence, for comparison we show this lower threshold line (given in red) in the figures. 

Although the amplitude spectrum (Figure 2) does not appear very complex, attempts to determine the underlying variations by prewhitening reveal that this star indeed has a complicated pulsation spectrum. 
After identifying the highest amplitude peak in the (computed) weighted power spectrum of the original data set, we went back to the unweighted FT to confirm the periodicity.  We then removed this peak from the data by subtracting a sine wave with the frequency, amplitude and phase determined by non-linear least-squares fit (NLLS) from the light curve.  In our NLLS fits, we use $T_{\rm max}$, the time of first maximum after $T_o$= 21 August 2004 0hUT, for the phase variable.  We calculate the power spectrum of the residuals (light curve of the entire data set with this sinusoid removed). Then we looked at this ``prewhitened'' weighted FT to find the next highest peak and repeated the procedure. We determined the values of frequencies step by step, carefully checking whether new frequencies affected the other values and
calculating a simultaneous $n$- frequency fit to the data until no new peaks could be identified with significance (with amplitude above $3.7\sigma_{\rm noise}$).

Figure 5 illustrates this procedure.  It shows the spectral window centered at 7000$\mu$Hz in the top panel, and the FT of the entire data set in the region around the highest amplitude peaks in the next panel. The FT, prewhitened by the two highest peaks, $f_{1}=7 088.67\mu$Hz with an amplitude of 3.0 mma and $f_{2}=6826.06\mu$Hz with an amplitude of 2.4 mma, is presented at the third panel from the top of Figure 5. The highest amplitude peak $f_{1}$ appears to have fine structure, since significant power remains above the 3.7$\sigma_{\rm noise}$ level (blue/top solid line) after removing the highest peak at 7088.67$\mu$Hz.  Two additional closely spaced peaks are present at frequencies $f_{3}=7091.0\mu$Hz and $f_{4}=7091.7\mu$Hz with amplitudes of 1.2 mma  and 1.1 mma , respectively.  Even though this frequency difference is formally unresolvable given our frequency resolution of $0.9\mu$Hz, both of these modes are easily resolved in our NLLS fit to the data. Furthermore, all the power above the 3$\sigma_{\rm noise}$ level (red/lower solid line) is gone after prewhitening the data with those three peaks, bottom panel of Figure 5.  The peak at $6826.06 \mu$Hz is cleanly removed by a single frequency prewhitening step.  

In the extended observations of sdBV stars it has been detected that the individual pulsation frequencies have amplitudes that vary in time. Such frequencies could introduce spurious peaks in the FT as both  the Fourier analysis and the prewhitening assume constant amplitude. Therefore we checked for the amplitude and the phase variability of the two highest amplitude periodicities ($f_{1}$ and $f_{2}$) as we were able to resolve them in each individual run. The peak at $7 088.67\mu$Hz appears to vary both in amplitude and in phase. The amplitude and phase of the $f_{2}$ mode does not appear to be changing throughout the run. The errors on amplitudes and phases for some runs are too high to allow us to make any claims. All that can be said is that there is a possibility of a periodic change in amplitude and phase for the $f_{1}$ frequency with a period of roughly five days, and that the frequency at $f_{2}$ appears to be stable both in amplitude and in phase over the run.  Figure 6 shows the behavior of the phase ($T_{\rm max}$) and amplitude of $f_{1}$ and $f_{2}$.  This figure shows the amplitude and phase of individual long runs; we grouped nearby short runs to reduce the phase and amplitude uncertainties when necessary.

The variability we see in $f_{1}$ can be the intrinsic amplitude variability of the mode or it can be caused by beating of two or more closely spaced frequencies. While the correct answer is impossible to give within the frame of the current time-series techniques, all the clues of our analysis point toward attributing the amplitude variability in $f_{1}$ to the beating of closely spaced frequencies.  If this was a stochastic amplitude variation, the signature in the Fourier transform would mimic two or more closely spaced modes of similar amplitude, and complete removal with a single sinusoid in the prewhitening process would not be appropriate.  Therefore, the beating of closely spaced frequencies is the most probable cause of the amplitude variability we see in $f_{1}$ as it is 
consistent with the fine structure of two additional closely spaced frequencies we found in the data. 

% Figure 5 
\begin{figure}
\plotone{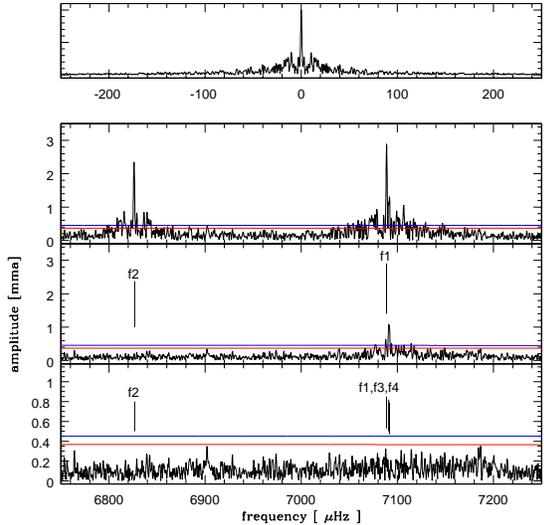}
\caption{Top panel: the spectral window centered at $7000\mu$Hz given on the same frequency scale as the data. Second panel: the FT of the entire WET data set on PG~0014+067 expanded around the region of the highest power. Third panel: the residual FT prewhitened by $f_{1}$ and $f_{2}$ at frequencies indicated.  Bottom panel: prewhitened FT by the first four frequencies from the Table 2. Note the different amplitude scale. }
\end{figure}

The next highest peak in the FT is at 7289.0$\mu$Hz.  Figure 7 displays this region of the frequency domain, showing the FT of the data before (top) and after (bottom) removing this periodicity, with the parameters determined by the NLLS fit to the data.  Although there is some power left above the 3$\sigma_{\rm noise}$ level (red/lower solid line) NLLS cannot satisfactorily converge on the parameters of these low amplitude peaks. Hence, prewhitening stops here for this region, but we do list the highest residual peak in this range as a possible real frequency in Table 3.

% Figure 6
\begin{figure}
\plotone{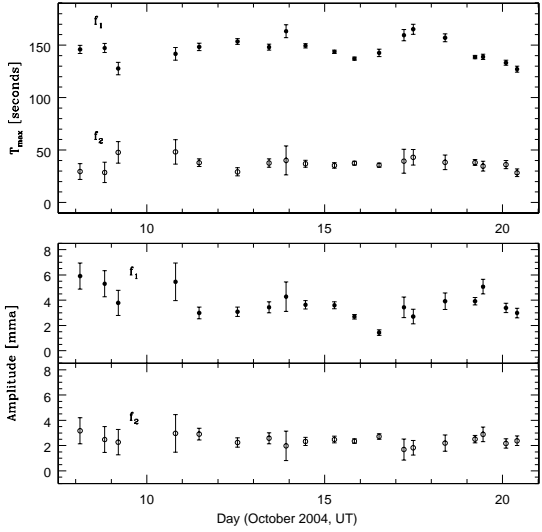}
\caption{Top panel: phase ($T_{\rm max}$, in seconds) as a function of time for the two principal oscillations in PG 0014+067.  Open circles are for $f_{2}$ and solid circles are for $f_{1}$.  Bottom panels: amplitude (in mma) for $f_{1}$ and $f_{2}$ over the course of the WET run.}
\end{figure}

% Figure 7 
\begin{figure}
\plotone{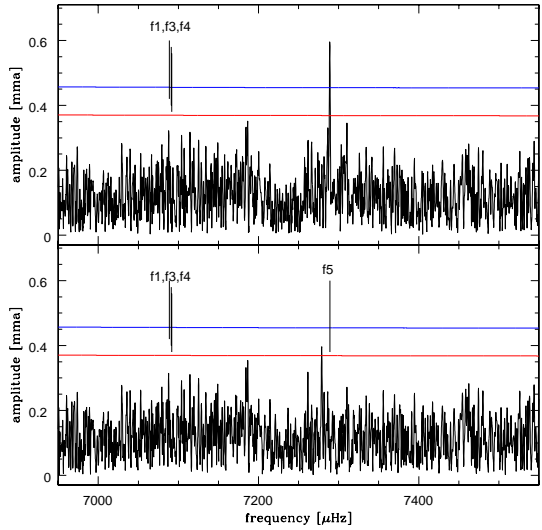}
\caption{Top panel:  the FT of the entire data set, pre-whitened by $f_1$ to $f_4$, expanded around $f_{5}$.
Bottom panel: the residual FT after prewhitening with  $f_{5}= 7289.0\mu$Hz.}
\end{figure}

Figure 8 shows the region of the next highest power; the top panel shows the residual FT after the five highest peaks have been removed ($f_{1}$ to $f_{5}$).  The highest power (amplitude) peak on the upper plot appears as a single peak of about 0.9 mma.  Detailed NLLS analysis finds 
two closely spaced modes at $f_{6}=6632.8\mu$Hz with an amplitude of 0.65 mma  and $f_{8}=6631.9\mu$Hz with an amplitude of 0.5 mma.  After those two sinusoids have been removed from the lightcurve, the residual spectrum stays clean, as it is shown on the middle panel of Figure 8.  The frequency difference between those two modes is of the order of our resolution, $0.9\mu$Hz.  The next highest peak is at $f_{9}=6452.9\mu$Hz with an amplitude just above our threshold (0.4 mma). Removal of this frequency reveals no more power above the noise (the bottom panel on Figure 8). The only power left in this region is at 6674$\mu$Hz, but at an amplitude that is below our adopted threshold, albeit above the 3$\sigma_{\rm noise}$ level.  We cannot confirm this frequency is real, we do however indicate this frequency in our ``almost'' list (Table 3).

% Figure 8 
\begin{figure}
\plotone{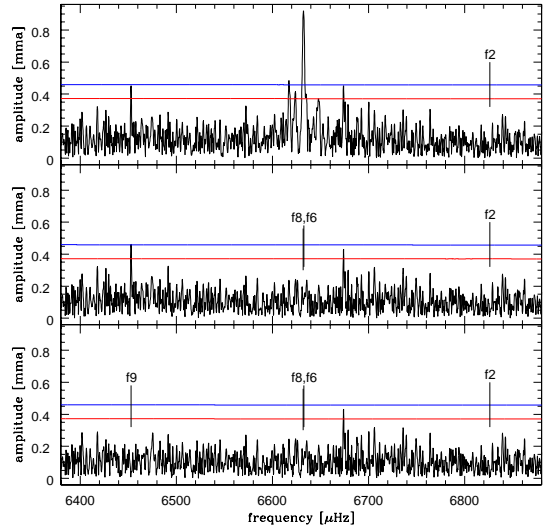}
\caption{Top panel: the residual FT after the first five modes have been removed from the data, in the range from 6380 to 6880 $\mu$Hz.
Middle panel: the temporal spectrum after $f_{6}=6 632.8\mu$Hz and $f_{8}=6631.9\mu$Hz have also been removed.  Bottom panel: the FT with additional prewhitening by the $f_{9}=6452.9\mu$Hz.}
\end{figure}

At the low frequency end, there is power with a rather low amplitude, at about $5923\mu$Hz.  Just by comparing with the spectral window (Figure 3) it is clear that we are most probably dealing with two or more closely spaced frequencies. NLLS finds a mode with an amplitude of 0.5 mma at $f_{7}=5 923.4\mu$Hz.  Figure 9 shows this region, giving the temporal spectrum before (top) and after this mode is removed by prewhitening (bottom panel). There is still excess power left after its removal, with a frequency of  $5921.3\mu$Hz.  NLLS is unable to converge on a simultaneous fit to these two modes along with the nine found previously. This second frequency is included into the list of possible true periodicities (Table 3).

% Figure 9 
\begin{figure}
\plotone{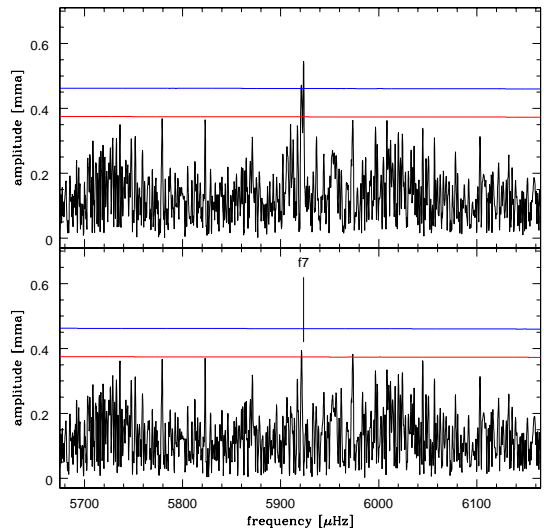}
\caption{The residual FT of the entire data set at the low frequency end of the main power (top) after $f_1$ through $f_6$, $f_8$, and $f_9$ are removed. Removal of $f_7$ results in the FT shown in the bottom frame.}
\end{figure}
The only power left above the threshold is just below 6200$\mu$Hz. The attempt to find the NLLS fit to the data including this mode was successful despite its low amplitude, and we have therefore removed this peak with a frequency of $f_{10}$=6193.5$\mu$Hz and an amplitude of 0.4 mma  from the total lightcurve. Prewhitening successfully removes all the excess power with only the noise left in that region.  Figure 10 shows the FT of the residuals. The top panel shows the starting spectrum, in this case the residual FT  prewhitened by the nine frequencies found, and the bottom is the FT after we have removed the ten frequencies.  

% Figure 10 
\begin{figure}
\plotone{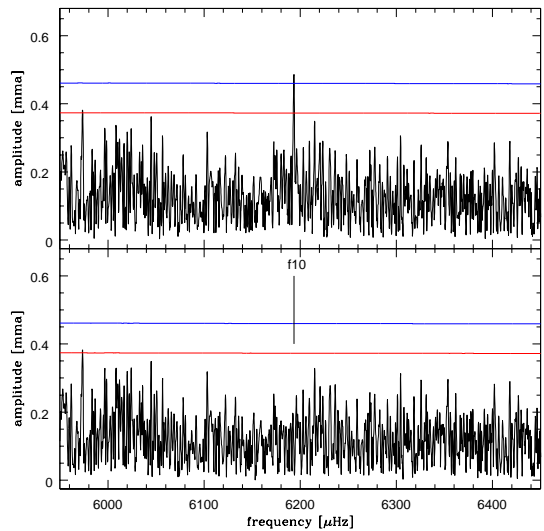}
\caption{FT near $f_{10}$ prewhitened by the nine frequencies, $f_{1}$ to $f_{9}$ (top) followed by residual FT after removing the tenth frequency $f_{10}$ (bottom).}
\end{figure}
% Table 2
\begin{deluxetable}{ccccc}
\tablewidth{0pt}
\tablecaption{Periodicities found in WET data on PG~0014+067.}
\tablehead{
\colhead{Mode}& \colhead{Frequency} & \colhead{Period} & \colhead{Amplitude} & \colhead{$T_{\rm max}$\tablenotemark{1}} \\ 
 & \colhead{[$\mu$Hz]} &  \colhead{[s]} &  \colhead{[mma]} & \colhead{[s]}
 }
\startdata
$f_{7}$         & 5923.4  $\pm$0.1& 168.821 $\pm$0.003& 0.54 $\pm$0.13& 20 $\pm$16 \\

$f_{10}$       & 6193.5  $\pm$0.2& 161.458 $\pm$0.005& 0.44 $\pm$0.13& 125 $\pm$20 \\
$f_{9}$         & 6452.9  $\pm$0.2& 154.968 $\pm$0.005 & 0.45 $\pm$0.13& 150 $\pm$18 \\
$f_{8}$         & 6631.9  $\pm$0.2& 150.786 $\pm$0.005 & 0.49 $\pm$0.15&   90 $\pm$20 \\
$f_{6}$         & 6632.8  $\pm$0.1& 150.766 $\pm$0.003 & 0.65 $\pm$0.15& 143 $\pm$15 \\

$f_{2}$         & 6826.06  $\pm$0.03& 146.4974 $\pm$0.0006 & 2.38 $\pm$0.13 & 28 $\pm$3\\
$f_{1}$         & 7088.67  $\pm$0.03& 141.0702 $\pm$0.0006 & 2.98 $\pm$0.13 & 8$\pm$3 \\
$f_{3}$         & 7091.0  $\pm$0.1& 141.023 $\pm$0.002& 1.22 $\pm$0.33& 133$\pm$11 \\
$f_{4}$         & 7091.7  $\pm$0.1& 141.011 $\pm0.002$& 1.10 $\pm$0.32& 55$\pm$11 \\

$f_{5}$         & 7289.0  $\pm$0.1& 137.193 $\pm$0.002& 0.65 $\pm$0.13&107$\pm$11 \\ \hline

\enddata
\tablenotetext{1}{time of first maximum after $T_o$= 21 August 2004, 0h UT}
\end{deluxetable}
To summarize, performing Fourier analysis and NLLS fitting on the entire WET data set obtained during  Xcov24, we have detected a total of ten frequencies in the power spectrum of  PG~0014+067 with a 99\% significance level and with a resolution of 0.9$\mu$Hz. 
The final results of a simultaneous multiple sine-wave fit to the total light curve are summarized in Table 2, listing the frequencies, amplitudes and phases of the main modes detected in the WET data set. The column labeled as \textbf{Mode} gives the name of each frequency as $f_n$ where $n$ is ordered by relative amplitude. The column labeled as \textbf{$T_{\rm max}$} is the time of the first maximum after August 21 2004 0h UT, given in seconds. The uncertainties for the values given in the table are formal least-square estimates of the 1$\sigma$ errors.

Figure 11 shows the FT of the combined WET data set on PG~0014+067 before (top) and after the removal (bottom) of ten periodicities listed in Table 2. We note that in the region where most peaks appear, the 3.7$\sigma_{\rm noise}$ level corresponds to about 0.44 mma.  There are thirteen low amplitude peaks below our adopted threshold (3.7$\sigma_{\rm noise}$) albeit above the somewhat less conservative acceptance criterion at 3$\sigma_{\rm noise}$. One of these, at 13,540 $\mu$Hz, lies outside of the range displayed in Figure 11.  These peaks are at best marginal detections. In Table 3 we list the frequencies, corresponding periods and amplitudes of those thirteen marginally detected modes should they be detected in other data sets.

% Figure 11 
\begin{figure}
\plotone{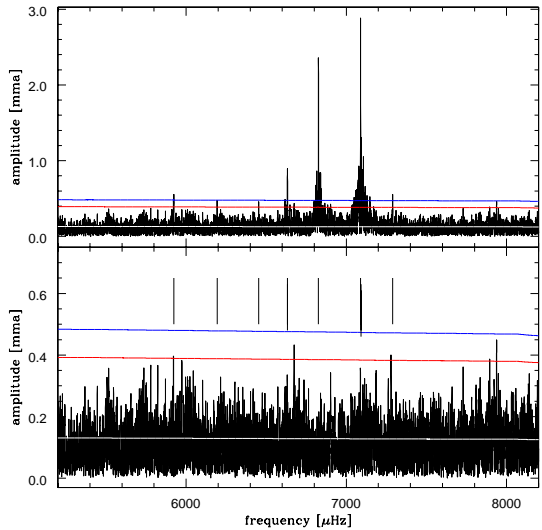}
\caption{Total FT of the original WET data set on PG~0014+067 (top) followed by the residual FT of the same data set, prewhitened by the ten frequencies listed in Table 2. Note that the red (lower) solid line corresponds to 3$\sigma_{\rm noise}$, while the blue (upper) solid line corresponds to 3.7$\sigma_{\rm noise}$. The white solid line is $\sigma_{\rm noise}$.}
\end{figure}
% Table 3
\begin{deluxetable}{ccc}
\tablecaption{Marginal detections- peaks with an amplitude between $3.7\sigma_{\rm noise}$ and $3\sigma_{\rm noise}$ seen in WET data on PG~0014+067.}
\tablehead{
\colhead{Frequency} & \colhead{Period} & \colhead{Amplitude} \\
\colhead{$[\mu$Hz$]$} & \colhead{ [s] } & \colhead{[mma]}
}
\tablewidth{0pt}
\startdata
 5921.3 & 168.882&  0.40 \\
 6674.0 & 149.835 &  0.43  \\
 7278.9 & 137.383 &  0.40  \\
 7895.0 & 126.662 & 0.39  \\
 7938.3 & 125.971 & 0.45  \\
 8870.8 & 112.729 & 0.42 \\
 9576.6 & 104.421 & 0.41  \\
 9834.2 & 101.686 & 0.38  \\
 10807.8 & 92.526 & 0.36 \\
 11116.5 & 89.956 & 0.36 \\
 11364.8 & 87.991 & 0.36 \\
 12083.7 & 82.756 &0.37  \\
 13539.8 & 73.856 &0.38\\
\enddata
\end{deluxetable}
\subsection{Comparing with previous data on PG~0014+067} 

In partial support of the run, six nights of observation of PG~0014+067 took place two months prior to Xcov24 (August  2004) using the high-speed multi-channel photometer \textsc{ultracam} 
\citep{ultracam2001} on the 4.2m William Herschel Telescope (WHT) at La Palma (Jeffery et al. 2005). We compare our results with published results on PG~0014+067 from  \citet{Brassard2001} and \citet{Jeffery2005}. Table 4 lists the periodicities identified by \citet{Jeffery2005} and \citet{Brassard2001} together with the WET results.

Note that all frequencies identified by the WET (Table 2) were identified by either Jeffery et al. ~(2005) or by  Brassard et al.~(2001) or by both to within the cycle per day aliases. The highest amplitude frequency in the WET data is at 7088.67$\mu$Hz and it is seen as a multiplet of three closely spaced frequencies (along with 7091.0$\mu$Hz and 7091.7$\mu$Hz). The \textsc{ultracam} data set identified 4 closely spaced frequencies in this range (see Table 4) out of which the highest amplitude one in their data set is at 7091.7$\mu$Hz. This multiplet is not identified in the CFHT data  probably due to the resolution of their run, only one frequency of this multiplet is found at 7088.7$\mu$Hz. The amplitude and phase of the highest amplitude mode (7088.67$\mu$Hz ) in the WET data set as well as the highest amplitude mode (7091.7$\mu$Hz) in the \textsc{ultracam} data set  were noticed to vary on the same time scale of five days. 

The amplitudes of the individual modes in sdBV stars are known to change on the time-scale of years, months and even days \citep{Kilkenny99}, although the question whether this results from true physical change in the star or beating between closely spaced frequencies still remains unanswered.  In this particular case both the WET and \textsc{ultracam} data sets find the amplitude and phase variability on the same time scale and identify closely spaced frequencies, a multiplet around 7090$\mu$Hz. This gives an additional credibility to the hypothesis that this relative amplitude change is caused by beating, though longer runs would be needed to confirm this speculation.

% Table 4 
\begin{deluxetable}{cccl}
\tabletypesize{\scriptsize}

\tablecaption{WET frequencies in PG~0014+067 (in $\mu$Hz) compared with the \textsc{ultracam} data, Jeffery et al.~(2005) and CFHT data, Brassard et al.~(2001). }
\tablehead{
\colhead{Mode} & \colhead{WET} & \colhead{ \textsc{ultracam}} & \colhead{CFHT}
 }
\tablewidth{0pt}
\startdata
\nodata & \nodata &  5780.9\tablenotemark{w} & \nodata \\
\nodata & \nodata & \nodata & 5896.2\\ 
$f_{7}$          & 5923.4 & 5921.9\tablenotemark{m} & 5923.2 \\
\nodata & \nodata  & 5924.8 & \nodata  \\
$f_{10}$       & 6193.5  &  \nodata  & 6227.7 =  \textbf{$f_{10}$} + 3d$^{-1}$\\
$f_{9}$         & 6452.9   & 6454.4  & \nodata  \\
$f_{8}$         & 6631.9  & \nodata   & 6630.7 \\
$f_{6}$         & 6632.8  & 6632.6 & 6621.1 =  \textbf{$f_{6}$} - 1d$^{-1}$ \\
\nodata  & \nodata  & 6646.5 & \nodata \\ 
\nodata  & \nodata  & 6659.9\tablenotemark{w} & \nodata  \\
\nodata  & \nodata  & 6726.8\tablenotemark{w} & \nodata \\ 
$f_{2}$         & 6826.06  & 6826.1  & 6837.5 = \textbf{$f_{2}$}+1d$^{-1}$\\
\nodata  & \nodata  & 7076.6\tablenotemark{w}= \textbf{$f_{1}$}-1d$^{-1}$ & 7079.1= \textbf{$f_{3}$} - 1d $^{-1}$\\
$f_{1}$  & 7088.67 & 7089.1\tablenotemark{m} & 7088.7\\
$f_{3}$         & 7091.0  & \nodata  &  \nodata \\
$f_{4}$         & 7091.7   & 7091.67\tablenotemark{*} &  \nodata \\
\nodata  & \nodata  & 7093.4 & \nodata \\
\nodata  & \nodata  & 7094.8 & \nodata \\
\nodata  & \nodata  & \nodata & 7150.2 \\
\nodata  & \nodata  & 7187.5 &  \nodata \\
$f_{5}$   & 7289.0  & \nodata  & 7286.2 \\ 
\nodata  & \nodata  & \nodata  & 7670.3 \\ 
\nodata  & \nodata  & \nodata  & 7952.1 \\ 
\nodata  & \nodata  & \nodata  & 8552.1 \\ 
\nodata  & \nodata  & 8588.9\tablenotemark{w} &  \nodata \\
\nodata  & \nodata  & \nodata  & 9797.6 \\
\nodata  & \nodata  & 9971.5 & 9970.3 \\
\nodata  & \nodata  & 11547.9 & \nodata \\
$(2 \times f_{10})$ & \nodata  & \nodata  & 12386.8 \\
$(2 \times f_9)$\tablenotemark{u} & \nodata  & 12910.9\tablenotemark{w} & \nodata  \\
\enddata
\tablenotetext{*}{highest amplitude seen in \textsc{ultracam} data}
\tablenotetext{w}{seen in white light only}
\tablenotetext{m}{multiplet}
\tablenotetext{u}{$f_9$ as seen in \textsc{ultracam} data}
\end{deluxetable}

\section{Discussion}

The frequencies we saw in PG~0014+067 during the October 2004 WET campaign showed no obvious patterns that could readily be attributed to rotational splitting or other common effects in nonradially pulsating stars.  However, some unusual systematics among the frequencies became readily apparent.  In this section we discuss those systematics, and explore the statistical significance of such patterns by including additional secure frequencies seen by \citet{Jeffery2005}.

\subsection{Strange systematics in the observed frequencies}

The frequency list of PG~0014+067 displays some suggestive systematics.  As an exercise in numerology, but one that will eventually provide some statistically significant results, we examined these systematics further.

Considering only the dominant modes in each closely--spaced multiplet, there are 7 main frequencies in the WET data on PG~0014+067, represented by $f_1$, $f_2$, $f_5$, $f_6$, $f_7$, $f_9$, and $f_{10}$.  Within this group, we find several pairs of frequencies separated by multiples of 90$\mu$Hz.  The first five lines in Table 5 list these differences.  While not all entries in the table are independent, it illustrates that some (large) frequency differences are very similar to one another, and all share a common factor of  $90.47 \pm 0.53$ $\mu$Hz.  

The frequency $f_5$ does not pair with any other mode in a difference that is a factor of 90.47 $\mu$Hz.  However, the average value of $f_5$ and $f_1$, 7188.8$\mu$Hz, is four times that interval from $f_2$.  This somewhat suspicious additional entry in Table 5 is supported by the observation of a frequency that lies within 1.3$\mu$Hz of that value by \citet{Jeffery2005} (the frequency of 7187.5$\mu$Hz in Table 4).  Thus the three frequencies ($f_5=7289.0\mu$Hz, 7187.5$\mu$Hz, and $f_1=7088.7 \mu$Hz) form a nearly equally-spaced triplet with a spacing of about 101.5$\mu$Hz.  In light of this, we explored the frequencies in Table 2 allowing for an $\approx 101 \mu$Hz deviation; these are shown in the last four lines of Table 5.  With this additional complication, additional evidence for a possible chain of frequencies separated by about 90$\mu$Hz appears.  If we adopt a value of 90.47$\mu$Hz for the first spacing, then the last four lines of Table 5 suggest that the second spacing is 101.1$\pm 1.4$ $\mu$Hz.

% Table 5
\begin{deluxetable}{rccl}
\tablecaption{Frequency differences in the PG~0014+067 WET data}
\tablehead{
\colhead{Frequency pair} & \colhead{difference} & \colhead{multiple}\\ 
    & \colhead{[$\mu$Hz]} & \colhead{[$\mu$Hz]} & }
\startdata
$f_{6} - f_{9}$ & 179.9 & = $2 \times 89.95$ \\
$f_{10} - f_{7}$ & 270.1 & = $3 \times 90.03$ \\
$f_{1} - f_{6}$ & 455.9 & = $5 \times 91.17$ \\
$f_{2} - f_{10}$ & 632.6 & = $7 \times 90.37$ \\
$f_{1} - f_{9}$ & 635.8 & = $7 \times 90.82$ & mean spacing = 90.47$\pm$ 0.53\\
$\langle f_{5},f_{1} \rangle - f_{2}$ & 362.8 & = $4 \times 90.69$ \\
\\
$f_{5} - f_{1}$ & 200.3 & = $2 \times 100.15$ \\
\\
$f_{1} - f_{2}$ & 262.6 & = $4 \times 90.47\tablenotemark{a} -\  99.29$ \\
$f_{5} - f_{2}$ & 462.9 & = $4 \times 90.47\tablenotemark{a} + 101.01$\\
$f_{9} - f_{10}$ & 259.4 & = $4 \times 90.47\tablenotemark{a} - 102.49 $\\
$f_{2} - f_{9}$ & 373.2 & = $3 \times 90.47\tablenotemark{a}  + 101.78 $
& mean spacing = 101.1$\pm$1.4\\
\\
\enddata
\tablenotetext{a}{fixed at mean value from lines 1-5}
\end{deluxetable}

We recognize that the spacings identified by this subjective technique alone may not be unique or even meaningful.  We note that there is no a priori expectation of a chain of equally spaced (in frequency) modes for these stars, nor is there an expectation, based on the physics of the star, for a splitting as large as 101$\mu$Hz.
Neither 90$\mu$Hz nor 100$\mu$Hz spacings are seen in theoretical models of these stars.  The asymptotic frequency spacing for $p-$modes in these stars (that is, modes with $n > l$, consecutive $n$, and alternating $l$) is roughly 750$\mu$Hz.  This value was determined using the model frequencies from \citet{Brassard2001} along with our own sdB models with appropriate values for $\log g$ and $T_{\rm eff}$.

Even so, could the 101$\mu$Hz splitting be a rotational splitting?  While this large splitting could be caused by rotation, the implied rotation rate (if solid-body rotation) although not ruled out, would be faster than any known 
single sdBV detected so far, based on spectroscopic study of line profiles.  Spacings this large are seen only in close binary pulsating sdBV stars such as PG~1336-018 \citep{Kilkenny2003}.  On the other hand \citet{KH2005} suggest that rapid internal rotation could produce large splittings in a star with a slow surface rotation rate, but their predictions suggest that the splittings of modes differing in $n$ and $l$ should not show the same value.

\subsection{An empirical, phenomenological relation}

With two apparent splittings present, we decided to explore an entirely phenomenological parameterization that could then be used to make an empirical fit to the observed frequencies, and provide a framework for determining the statistical significance of these patterns.  We chose a form reminiscent of asymptotic $p-$mode pulsation with a constant rotation frequency 
\begin{equation}
f(i,j) = f_o + i \times \delta + j \times \Delta
\end{equation}
where $\delta$ represents a small spacing (and $i$ is an integer ranging from 0 upwards) and $\Delta$ represents a large spacing (with $j$ initially limited to being either -1, 0, or 1).  In the equation above, $f_o$ represents a zero-point for the fit with $i=j=0$.

In the general case of fitting an observed set of frequencies, we performed a two-dimensional $\chi^2$ minimization over a grid of values of $\delta$ from 40 to 120 $\mu$Hz and $\Delta$ from 80 to 210 $\mu$Hz.  For each ($\delta$, $\Delta$) pair, we found the combination of ($i$, $j$) that minimized the root-mean-square (RMS) difference between the observed and model frequencies, and computed the $\chi^2$ of the fit.  Clearly, there is an aliasing problem when the combinations of $i$, $j$, $\delta$, and $\Delta$ produce commensurate spacings, so the $\chi^2$ surface shows multiple minima.  To help break that degeneracy, we impose an additional criterion that a $(\delta, \Delta)$ solution result in at least two modes that have the same value of $i$ but different $j$.  The quantized nature of the problem (the values of $i$ and $j$ are constrained to be integers) and the additional constraint on the values of $j$ complicate determination of statistical significance of any fits.  Therefore to find the form of the distribution of $\chi^2$ for a random selection of frequencies, we performed a series of Monte Carlo trials using frequency lists drawn from uniformly random distributions within the range shown by PG~0014+067.

Using only the WET frequencies, we find two solutions of high significance; however, only one satisfies the requirement of having two modes with the same value of $i$ but different values of $j$.  That solution is the one found by the successive differencing procedure described in the previous subsection.  The fit values were $\delta =90.37$ $\mu$Hz, $\Delta = 101.46$ $\mu$Hz, and $f_o$=5922.46 $\mu$Hz.

With only seven frequencies and more than three free parameters (the values of $\delta$, $\Delta$ and $f_o$ are free, but the choices of $i$ and $j$ are constrained), the fit above is not well constrained.  Via the Monte Carlo technique with 7 frequencies, we find the fit to be significant at the 95\% confidence level; that is, in a trial of 285 frequency sets, only 13 had a lower value of $\chi^2$ and at least two modes with the same value of $i$ and different values of $j$.   While this is encouraging in terms of suspecting that the parameterization above may be meaningful, it is not, by itself, very convincing.

Fortunately, we have additional frequencies in this star that have been reliably measured by others that lie within the frequency range from the WET data alone.
This list is given in Table 6.  By ``reliable'' we mean peaks seen in more than one investigation, and those that are clearly true peaks and not aliases.  Given the exceptional quality and clean window of the \textsc{ultracam} data, we give that data set higher relative weight, using the WET to resolve residual ambiguities originating from diurnal aliases.  Using these additional observations, five additional frequencies in the range that the WET showed pulsations can be examined for compliance with the relation derived using only the seven WET frequencies in Table 4.  

Of the five additional frequencies in Table 6, four fit the pattern determined from the WET frequencies alone \textit{using the same parameters}.  Again, we use Monte Carlo simulations of this process to judge whether or not this is a significant effect.  In this case we measure the RMS deviation of five additional randomly chosen frequencies from the previously--established relation from the WET data. The RMS deviation is calculated after dropping the worst-fitting of the five.  Comparing the distribution of the RMS deviations in the Monte Carlo trials with that from the data in Table 6 reveals that the fit of the relation (equation 1) with the data is significant at the 99.4\% confidence level (based on 9829 trials) under these conditions.

In practice, we impose an additional requirement that the fit must include modes that have at least two pairs with the same $i$ but different $j$ (i.e. that at least two values of $i$ show more than one $j$ component). 
Enforcing this additional condition on the fit criteria restrict the aliasing problem and result in a smaller number of possible solutions.  This process also results in a much higher statistical significance for the fit to PG~0014+067; the confidence level rises to the 99.95\% level. In the fit in Table 6, we see that there are three values of $i$ that have more than one $j$ component, with one complete `triplet' - a concordance that none of the 9829 Monte Carlo trials reached.

Finally, we can employ the fit procedure using all of the modes in Table 6 to determine more accurate values of the parameters of Equation 1. For PG~0014+067 this procedure yields values of $\delta$=90.37, $\Delta$=101.22, and $f_o$=5923.24 (all in $\mu$Hz).  Using these values in Equation 1, we find the model frequencies listed in Table 6.  Note that the fit is extremely good - all but one of the 12 modes identified in the table are fit to within 0.05\%, with an RMS difference of 0.013~$\delta$.  To illustrate the closeness of the fit, Figure 12 shows the frequencies of PG~0014+067 in an ``echelle diagram'' similar to those used in helioseismology.  The vertical axis is frequency, and the horizontal axis (also in frequency units) shows the departure of each mode from an integral multiple of the small spacing $\delta$.  The repeating pattern of $j=0$ modes stacks above an ordinate of 0, while modes with nonzero values of $j$ lie on either side at frequencies $\Delta$ away.

% Table 6
\begin{deluxetable}{lccccccc}
\tablecaption{Pulsation frequencies in PG~0014+067, along with an empirical model fit.}
\tablehead{
\colhead{\#} & \colhead{Frequency} & \colhead{Amplitude} & \colhead{note} & \colhead{$i$} & \colhead{$j$} & \colhead{Model} & \colhead{Difference}\\ 
     & \colhead{[$\mu$Hz]} & \colhead{[mma]} & & & & \colhead{[$\mu$Hz]} & \colhead{[$\mu$Hz]}
 }
\startdata
\nodata & 5780.9 & 0.35 & \textsc{ultracam} (w) & \nodata & \nodata & 
\nodata & \nodata \\
$f_{7}$  & 5923.4 & 0.54 & fine structure & 0 & 0 & 5923.2 & 0.2 \\
$f_{10}$  & 6193.5 & 0.44 &  & 3 & 0 & 6194.1 & -0.6\\
$f_{9}$  & 6452.9 & 0.45 &  & 7 & -1 & 6454.1 & -1.2 \\
$f_{6}$ ($f_{8}$)  & 6632.8 & 0.65 & fine structure & 9 & -1 & 6634.6 & -1.8 \\
\nodata & 6646.5 & 0.60 & \textsc{ultracam} & 8 & 0 & 6645.6 & 0.9\\
\nodata  & 6659.9 & 0.34 & \textsc{ultracam} (w) & 7 & 1 & 6656.5 & -3.4 \\
\nodata  & 6726.8 & 0.37 & \textsc{ultracam} (w) & 10 & -1 & 6724.9 & 1.9 \\
$f_{2}$  & 6826.1 & 2.38 &  & 10 & 0 & 6826.1 & 0.0 \\
$f_{1}$ ($f_{3}, f_{4}$) & 7088.7 & 2.98 & fine structure & 14 & -1 & 7086.1 & 2.6\\
- & 7187.5 & 0.66 & \textsc{ultracam} & 14 & 0 & 7187.3 & 0.2\\
$f_{5}$ & 7289.0 & 0.65 &  & 14 & 1 & 7288.5 & 0.5\\
\enddata
\end{deluxetable}

% Figure 12 
\begin{figure}
\plotone{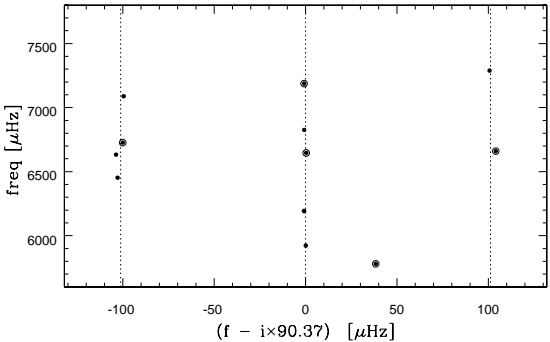}
\caption{``Echelle'' diagram of the frequencies in PG~0014+067; the frequencies have been folded on the small spacing $\delta = 90.37$ and stacked, showing the uniformity of the frequency spacings within the main band and the two side bands separated from the main band by the large spacing $\Delta = 101.22$.  Expanded points are from the non-WET modes in Table 6.}
\end{figure}

Is this asymptotic pulsation? After all, high--order $p$-modes show a more-or-less constant frequency spacing; such sequences are seen in helioseismic data and in the rapidly oscillating Ap stars (i.e. \citet{Kurtz2005}).
The sequence of modes split by integral multiples of $\delta$ cannot be asymptotic $p-$mode behavior.  As mentioned above, models of PG~0014+067, and sdBV pulsators in general, indicate that the radial fundamental frequency in the models is usually close (in frequency) to the observed mode frequencies.  Asymptotic relations such as Equation (1) are usually valid (at the few-percent level) only for values of $n \gg l$, or more generally for large values of $n$.  Even so, the computed frequency separation for $p-$modes in sdBV models yields values of several hundred $\mu$Hz - a factor of 5 or more larger than what PG~0014+067 shows.

In summary, the frequencies seen in PG~0014+067 obey a simple empirical trend - one that cannot be understood in terms of standard evolutionary and pulsation models for such stars.  Simply put, there is no applicable physics in this purely empirical fit.  In a subsequent paper, we will examine the frequencies present in other well--studied sdB stars, and we show that they show a similar pattern as well.  The theoretical interpretation of this pattern is not at all clear, but we anticipate that the systematic frequency differences discovered in PG~0014+067 and other sdB stars could be a very important clue to unlocking some secrets of their interiors.

\acknowledgments

Financial support for this work came from the U.S. National Science Foundation, through grant AST20205983 to Iowa State University. As of 1 September 2005, MV is a PhD student of the Research Council of 
the KULeuven under the doctoral scholarship OE/05/20. She acknowledges additional financial support from KULeuven grant 
GOA/2003/04. MV would like to thank Elizabeth Potter and Chris Tourek for helping the Xcov24 headquarters and with the data reduction. ZC and MP acknowledge financial support from Hungarian OTKA T-038440 and T-046207. NSF grant AST037480 to Missouri State University supported traveling costs and the camera used by MDR.
SJOT is supported by the Deutsches Zentrum fŸr Luft- und Raumfahrt Ê(DLR) through grant no. 50-OR-0202.

\end{document}